\RequirePackage{lineno}

\documentclass[aps, prd, preprint, twocolumn, superscriptaddress, showpacs, amsmath, amssymb, tightenlines, nofootinbib, 10pt]{revtex4-1}

\usepackage{graphicx} 
\usepackage{dcolumn}  
\usepackage{bm}
\usepackage{hyperref}
\usepackage[caption=false]{subfig}

\usepackage[absolute]{textpos}

\newcolumntype{"}{@{\hskip\tabcolsep\vrule width 1pt\hskip\tabcolsep}}

\newcommand*\patchAmsMathEnvironmentForLineno[1]{%
  \expandafter\let\csname old#1\expandafter\endcsname\csname #1\endcsname
  \expandafter\let\csname oldend#1\expandafter\endcsname\csname end#1\endcsname
  \renewenvironment{#1}%
     {\linenomath\csname old#1\endcsname}%
     {\csname oldend#1\endcsname\endlinenomath}}%
\newcommand*\patchBothAmsMathEnvironmentsForLineno[1]{%
  \patchAmsMathEnvironmentForLineno{#1}%
  \patchAmsMathEnvironmentForLineno{#1*}}%
\AtBeginDocument{%
\patchBothAmsMathEnvironmentsForLineno{equation}%
\patchBothAmsMathEnvironmentsForLineno{align}%
\patchBothAmsMathEnvironmentsForLineno{flalign}%
\patchBothAmsMathEnvironmentsForLineno{alignat}%
\patchBothAmsMathEnvironmentsForLineno{gather}%
\patchBothAmsMathEnvironmentsForLineno{multline}%
}

\newcommand{\mbc}{M_{\textrm{bc}}}
\newcommand{\bsig}{B_{\textrm{sig}}}
\newcommand{\btag}{B_{\textrm{tag}}}
\newcommand{\gsig}{E_{\gamma}^{\textrm{sig}}}

\newcommand{\pnusig}{|\vec{\,p}_{\nu}^{\textrm{\,sig}}|}
\newcommand{\lb}{\lambda_B}
\newcommand{\ebeam}{E_{\textrm{beam}}}
\newcommand{\qq}{q\bar{q}}

\newcommand{\miss}{m^2_{\textrm{miss}}}
\newcommand{\bbar}{B \bar{B}}

\newcommand{\Bmunu}{B^+ \to \mu^+ \nu_\mu \gamma}
\newcommand{\Benu}{B^+ \to e^+ \nu_e \gamma}
\newcommand{\Blnu}{B^+ \to \ell^+ \nu_\ell \gamma}
\newcommand{\Blep}{B^+ \to \ell^+ \nu_\ell}
\newcommand{\Bpilnu}{B^+ \to \ell^+ \nu_\ell \pi^0}
\newcommand{\Betalnu}{B^+ \to \ell^+ \nu_\ell \eta}
\newcommand{\BXlnu}{B \to X_u \ell^+ \nu_\ell}
\newcommand{\bulnu}{b \to u \ell^+ \nu_\ell}
\newcommand{\coslg}{\cos \Theta_{\gamma\ell}}
\newcommand{\cosgnu}{\cos \Theta_{\gamma\nu}}

\newcommand\BRelimit{6.1}
\newcommand\BRmulimit{3.4}
\newcommand\BRllimit{3.5}
\newcommand\lambdalimit{238}
\newcommand\lambdaulimit{410}
\newcommand\lambdallimit{172}

\newcommand{\Journal}[4]{{#1} {\bf #2}, #3 (#4)}

\begin{document}

%
%
%
\preprint{\vbox{ \hbox{   }
						 \hbox{\textbf  {Belle Preprint  2015-6}}	
						 \hbox{\textbf {KEK Preprint  2015-3}}	   		       
}}

\title{\quad\\[1.0cm] \boldmath Search for $\Blnu$ decays with hadronic tagging using the full Belle data sample}

\noaffiliation
\affiliation{University of the Basque Country UPV/EHU, 48080 Bilbao}
\affiliation{Beihang University, Beijing 100191}
\affiliation{University of Bonn, 53115 Bonn}
\affiliation{Budker Institute of Nuclear Physics SB RAS and Novosibirsk State University, Novosibirsk 630090}
\affiliation{Faculty of Mathematics and Physics, Charles University, 121 16 Prague}
\affiliation{University of Cincinnati, Cincinnati, Ohio 45221}
\affiliation{Deutsches Elektronen--Synchrotron, 22607 Hamburg}
\affiliation{Justus-Liebig-Universit\"at Gie\ss{}en, 35392 Gie\ss{}en}
\affiliation{SOKENDAI (The Graduate University for Advanced Studies), Hayama 240-0193}
\affiliation{Hanyang University, Seoul 133-791}
\affiliation{University of Hawaii, Honolulu, Hawaii 96822}
\affiliation{High Energy Accelerator Research Organization (KEK), Tsukuba 305-0801}
\affiliation{IKERBASQUE, Basque Foundation for Science, 48013 Bilbao}
\affiliation{Indian Institute of Technology Guwahati, Assam 781039}
\affiliation{Indian Institute of Technology Madras, Chennai 600036}
\affiliation{Indiana University, Bloomington, Indiana 47408}
\affiliation{Institute of High Energy Physics, Vienna 1050}
\affiliation{Institute for High Energy Physics, Protvino 142281}
\affiliation{INFN - Sezione di Torino, 10125 Torino}
\affiliation{Institute for Theoretical and Experimental Physics, Moscow 117218}
\affiliation{J. Stefan Institute, 1000 Ljubljana}
\affiliation{Kanagawa University, Yokohama 221-8686}
\affiliation{Institut f\"ur Experimentelle Kernphysik, Karlsruher Institut f\"ur Technologie, 76131 Karlsruhe}
\affiliation{Kennesaw State University, Kennesaw GA 30144}
\affiliation{King Abdulaziz City for Science and Technology, Riyadh 11442}
\affiliation{Korea Institute of Science and Technology Information, Daejeon 305-806}
\affiliation{Korea University, Seoul 136-713}
\affiliation{Kyungpook National University, Daegu 702-701}
\affiliation{\'Ecole Polytechnique F\'ed\'erale de Lausanne (EPFL), Lausanne 1015}
\affiliation{Faculty of Mathematics and Physics, University of Ljubljana, 1000 Ljubljana}
\affiliation{Luther College, Decorah, Iowa 52101}
\affiliation{University of Maribor, 2000 Maribor}
\affiliation{Max-Planck-Institut f\"ur Physik, 80805 M\"unchen}
\affiliation{School of Physics, University of Melbourne, Victoria 3010}
\affiliation{Moscow Physical Engineering Institute, Moscow 115409}
\affiliation{Moscow Institute of Physics and Technology, Moscow Region 141700}
\affiliation{Graduate School of Science, Nagoya University, Nagoya 464-8602}
\affiliation{Kobayashi-Maskawa Institute, Nagoya University, Nagoya 464-8602}
\affiliation{Nara Women's University, Nara 630-8506}
\affiliation{National Central University, Chung-li 32054}
\affiliation{Department of Physics, National Taiwan University, Taipei 10617}
\affiliation{H. Niewodniczanski Institute of Nuclear Physics, Krakow 31-342}
\affiliation{Nippon Dental University, Niigata 951-8580}
\affiliation{Niigata University, Niigata 950-2181}
\affiliation{Osaka City University, Osaka 558-8585}
\affiliation{Pacific Northwest National Laboratory, Richland, Washington 99352}
\affiliation{Peking University, Beijing 100871}
\affiliation{University of Pittsburgh, Pittsburgh, Pennsylvania 15260}
\affiliation{University of Science and Technology of China, Hefei 230026}
\affiliation{Seoul National University, Seoul 151-742}
\affiliation{Soongsil University, Seoul 156-743}
\affiliation{University of South Carolina, Columbia, South Carolina 29208}
\affiliation{Sungkyunkwan University, Suwon 440-746}
\affiliation{School of Physics, University of Sydney, NSW 2006}
\affiliation{Department of Physics, Faculty of Science, University of Tabuk, Tabuk 71451}
\affiliation{Tata Institute of Fundamental Research, Mumbai 400005}
\affiliation{Excellence Cluster Universe, Technische Universit\"at M\"unchen, 85748 Garching}
\affiliation{Toho University, Funabashi 274-8510}
\affiliation{Tohoku University, Sendai 980-8578}
\affiliation{Department of Physics, University of Tokyo, Tokyo 113-0033}
\affiliation{Tokyo Institute of Technology, Tokyo 152-8550}
\affiliation{Tokyo Metropolitan University, Tokyo 192-0397}
\affiliation{University of Torino, 10124 Torino}
\affiliation{CNP, Virginia Polytechnic Institute and State University, Blacksburg, Virginia 24061}
\affiliation{Wayne State University, Detroit, Michigan 48202}
\affiliation{Yamagata University, Yamagata 990-8560}
\affiliation{Yonsei University, Seoul 120-749}

 \author{A.~Heller}\affiliation{Institut f\"ur Experimentelle Kernphysik, Karlsruher Institut f\"ur Technologie, 76131 Karlsruhe} 
 \author{P.~Goldenzweig}\affiliation{Institut f\"ur Experimentelle Kernphysik, Karlsruher Institut f\"ur Technologie, 76131 Karlsruhe} 
 \author{M.~Heck}\affiliation{Institut f\"ur Experimentelle Kernphysik, Karlsruher Institut f\"ur Technologie, 76131 Karlsruhe} 
 \author{T.~Kuhr}\affiliation{Institut f\"ur Experimentelle Kernphysik, Karlsruher Institut f\"ur Technologie, 76131 Karlsruhe} 
 \author{A.~Zupanc}\affiliation{J. Stefan Institute, 1000 Ljubljana} 
  \author{A.~Abdesselam}\affiliation{Department of Physics, Faculty of Science, University of Tabuk, Tabuk 71451} 
  \author{I.~Adachi}\affiliation{High Energy Accelerator Research Organization (KEK), Tsukuba 305-0801}\affiliation{SOKENDAI (The Graduate University for Advanced Studies), Hayama 240-0193} 
  \author{K.~Adamczyk}\affiliation{H. Niewodniczanski Institute of Nuclear Physics, Krakow 31-342} 
  \author{H.~Aihara}\affiliation{Department of Physics, University of Tokyo, Tokyo 113-0033} 
  \author{K.~Arinstein}\affiliation{Budker Institute of Nuclear Physics SB RAS and Novosibirsk State University, Novosibirsk 630090} 
% ...
\collaboration{The Belle Collaboration}

\begin{abstract}
We search for the decay $\Blnu$ with $\ell^+ = e^+$ or $\mu^+$ using the full Belle data set of $772\,\times 10^6 \bbar$ pairs, collected at the $\Upsilon(4\textrm{S})$ resonance with the Belle detector at the KEKB asymmetric-energy $e^+e^-$ collider. We reconstruct one $B$ meson in a hadronic decay mode and search for the $\Blnu$ decay in the remainder of the event. We observe no significant signal within the phase space of $\gsig > 1$\,GeV and obtain upper limits of ${\cal B}(\Benu) < \BRelimit \times 10^{-6}$, ${\cal B}(\Bmunu) < \BRmulimit \times 10^{-6}$, and ${\cal B}(\Blnu) < \BRllimit \times 10^{-6}$ at 90\% credibility level. 
\end{abstract}


\maketitle


{\renewcommand{\thefootnote}{\fnsymbol{footnote}}}
\setcounter{footnote}{0}

\section{Introduction}
The semileptonic decay $\Blnu$~\cite{CC} proceeds via a $\bar{b}u$ annihilation into a $W^+$ boson that decays into a lepton-neutrino pair. This is accompanied by a photon emission from one of the participating charged particles with emission from the up quark being the dominant contribution. The decay can be computed in Heavy Quark Effective Theory~\cite{beneke11}, which is valid for a high energetic photon emission above the QCD scale of ${E_{\gamma} \gg \Lambda_{QCD}}$. The resulting decay amplitude depends on the first inverse moment ${\lb^{-1}=\int_{0}^{\infty} d\omega \Phi_{B^{+}}(\omega)/\omega}$, where ${\Phi_{B^{+}}(\omega)}$ is the $B$ meson light-cone distribution amplitude in the high energy limit. This parameter is an important input to the QCD factorization scheme used in non-leptonic $B$ decay amplitudes~\cite{beneke99}; a tighter limit on --- or, {\it a fortiori}, a measurement of $\lb$ would improve the predictions for all of these processes. To produce consistent results for color-suppressed modes in non-leptonic $B$ decays, values of roughly ${\lb \approx 200\,\textrm{MeV}}$ are needed. The parameter cannot be calculated reliably by theory and thus has to be measured experimentally. The decay discussed in this Letter is advantageous since no additional unknown parameters are needed for its calculation in leading order.
\par
The branching fraction of the decay $\Blnu$ is expected to be larger than that of the purely leptonic $\Blep$ decay as the photon removes the helicity suppression of the process, thus enhancing the weak decay amplitude. This effect is diminished by the additional electromagnetic coupling introduced by the photon emission. The $\Blnu$ decay has been calculated up to first-order corrections in $1/m_b$ and radiative corrections at next-to-leading logarithmic order~\cite{beneke11}. The differential branching fraction is given by
\begin{equation}
 \frac{d\Gamma}{dE_{\gamma}}=\frac{\alpha_{em}G_{F}^{2}\lvert V_{ub}\rvert^{2}}{48\pi^{2}}m_{B}^{4}(1-x_{\gamma})x_{\gamma}^{3}\Big[F_{A}^{2} + F_{V}^{2}\Big], \label{eq:singlediff}
\end{equation}
with $x_{\gamma}=2E_{\gamma}/m_{B}$. Here, $m_{B}$ is the $B$ meson mass, $G_F$ the Fermi coupling constant, $V_{ub}$ the CKM matrix element, and $F_{A}$ and $F_{V}$ the axial and vector form factors, respectively. The form factors are given by 
\begin{eqnarray*}
     F_{V}(E_{\gamma}) & = &  \frac{Q_{u}m_{B}f_{B}}{2E_{\gamma}\lb(\mu)}R(E_{\gamma},\mu) \\
          & + & \Big[\xi(E_{\gamma})+\frac{Q_{u}m_{B}f_{B}}{(2E_{\gamma})^{2}}+\frac{Q_{b}m_{B}f_{B}}{2E_{\gamma}m_{b}}\Big],  \\
  \\
F_{A}(E_{\gamma}) & = &  \frac{Q_{u}m_{B}f_{B}}{2E_{\gamma}\lb(\mu)}R(E_{\gamma},\mu) \\
          & + & \Big[\xi(E_{\gamma})-\frac{Q_{u}m_{B}f_{B}}{(2E_{\gamma})^{2}}-\frac{Q_{b}m_{B}f_{B}}{2E_{\gamma}m_{b}}+\frac{Q_{\ell}f_{B}}{E_{\gamma}}\Big],
  \end{eqnarray*}
where $Q_{\ell, u, b}$ are the charges of the lepton, up quark, and bottom quark, respectively, $f_B$ is the decay constant for the $B$ meson, and $R(E_{\gamma},\mu)$ is the radiative correction calculated at the energy scale $\mu$. The first term in the form factors containing $\lb$ represents the leading-order contribution of the QCD heavy-quark expansion describing the photon emission by the light quark. The leading order term is corrected for higher-order radiative effects, with the $R(E_{\gamma},\mu)$ factor containing mass corrections for the up quark. The remaining terms in square brackets are $1/m_{b}$ power corrections which are: higher-order contributions for the hard and soft photon emission of the up quark ($Q_{u}$ and the ${\xi(E_{\gamma})\textrm{-term}}$, respectively); the photon emission by the $b$ quark, which is suppressed due to its higher mass (${Q_{b}\textrm{-term}}$); and the photon emission by the lepton, which is only present in the axial form factor (${Q_{\ell}\textrm{-term}}$). The radiative corrections contained in $R(E_{\gamma},\mu)$ reduce the leading-order amplitude by about $20-25\%$. The remaining $1/m_{b}$ power corrections have considerable parametric uncertainties. However, using central values for the parameters the power-suppressed terms reduce the decay amplitude by about half the amount of the radiative corrections. The soft correction for the light quark $\xi(E_{\gamma})$ constitutes the largest uncertainty in the form factors and it has been calculated in Ref.~\cite{braun13} to a higher precision.
\par
The most stringent limits for the decay process have been reported by the BaBar collaboration~\cite{aubert09} at 90\% confidence level with ${{\cal B}(\Benu) < 17 \times 10^{-6}}$, ${{\cal B}(\Bmunu) < 26 \times 10^{-6}}$, ${\cal B}(\Blnu) <15.6 \times 10^{-6}$, and a partial branching fraction ${\Delta{\cal B}(\Blnu) <14 \times 10^{-6}}$ for photons with energies higher than 1\,GeV. For the preferred value of ${\lb \approx 200\,\textrm{MeV}}$, a Standard Model branching fraction of $ {{\cal B}(\Blnu) \approx\,{\cal{O}}(10^{-6})}$ is expected~\cite{beneke11}. 

\section{Data sample and simulation}
This study uses a sample of $(771.6\pm 10.6)\times 10^6$ $B\bar{B}$ pairs, which corresponds to an integrated luminosity of ${711\,{\textrm {fb}^{-1}}}$ collected with the Belle detector at the KEKB asymmetric-energy $e^+e^-$ collider~\cite{kurokawa03}. The collider operates at the $\Upsilon(4\textrm{S})$ resonance with a center-of-mass energy of 10.58\,GeV/$c^2$, where the resonance decays almost exclusively to $\bbar$ pairs. 
\par
The Belle detector is a large-solid-angle magnetic spectrometer that consists of a silicon vertex detector, a 50-layer central drift chamber (CDC), an array of aerogel threshold Cherenkov counters (ACC), a barrel-like arrangement of time-of-flight scintillation counters (TOF), and an electromagnetic calorimeter (ECL) comprising CsI crystals located inside a superconducting solenoid coil that provides a 1.5 T magnetic field. An iron flux return located outside the coil is instrumented to detect $K^0_{\textrm L}$ mesons and to identify muons (KLM). A detailed description of the Belle detector can be found in Ref.~\cite{abashian02}.
\par
The analysis procedure is determined using Monte Carlo (MC) samples that are simulated with the EvtGen software package~\cite{lange01} followed by detector simulation performed with GEANT3~\cite{brun84}. Beam background is recorded by the experiment and added to each event in the simulated MC. Samples of ${2\times10^6}$ events are generated for each signal MC channel, where the latest theoretical calculation~\cite{beneke11} is implemented as a decay model in EvtGen. Different samples with high integrated luminosity are used to estimate the background. A MC sample containing resonant charmed $\bbar$ events with $b\to c$ decays contains ten times the integrated luminosity of the data sample. Non-resonant $e^+e^- \to \qq\,(q = u, d, s, c)$ continuum processes are included in a MC sample with six times the integrated luminosity of the data sample. A semileptonic ${\bulnu}$ sample with 20 times the statistics of the data contains the important background processes of $\Bpilnu$ and $\Betalnu$. For the latter two processes, high statistics MC is produced with about 100 times the size of the data sample. A final sample contains rare ${b\to s}$ transitions and additional processes with 50 times the integrated data luminosity. 
\section{Hadronic $B$-tagging}
As the neutrino of the signal decay cannot be detected, the full reconstruction technique provides strong constraints on the kinematics of the signal decay. The hadronic full reconstruction at Belle is a hierarchical reconstruction scheme of one of the two $B$ mesons (tag-side $\btag$ meson)~\cite{feindt11} in the event. 
\par
The charged $\btag$ meson candidate is reconstructed in one of 17 final states: $\bar{D}^{(*)} X_{\textrm had}$ (7 states), $\bar{D}^{(*)0} D_s^{(*)+}$ (4 states), $\bar{D}^0 K^+$, $D^-\pi^+\pi^+$, $J/\psi K^+ (\pi^0,\; \pi^+\pi^-)$, and $J/\psi K_S^0 \pi^+$, where $X_{\textrm had}$ is a set of selected states with one to four pions, of which one can be neutral. The $J/\psi$ particles are reconstructed from $e^+e^-$ or $\mu^+\mu^-$ decays. Two charged tracks are used to reconstruct a $K_S^0$ candidate whose mass must be within a 30\,MeV/$c^2$ window around the nominal $K_S^0$ mass. Neutral pions are reconstructed from pairs of photons, each with an energy of at least 30\,MeV and an invariant mass within 19\,MeV/$c^2$ of the nominal pion mass. Photons are identified as energy depositions in the calorimeter above 20\,MeV without an associated track. Charged tracks are identified as pions or kaons using a likelihood ratio constructed from CDC, ACC, and TOF information. Charged-track quality is improved by requiring that $|dz| < 4.0\,$cm and $dr < 2.0\,$cm, where $|dz|$ and $dr$ are the distances of closest approach of the track to the interaction point along the beam axis and in the transverse plane, respectively.
\par
The efficiency of the $\btag$ full reconstruction depends on the complexity of the decay of the signal-side $B$ meson. For the simple $\Blnu$ process, a relatively high efficiency of 0.6\% is found in the signal MC for correctly reconstructed $\btag$ candidates; for $b\to c$ processes, the efficiency lies around 0.2\%. 
\par
The full reconstruction contains a separate neural network (NN) for each particle type and decay mode and is trained with the NeuroBayes software~\cite{feindt06}. Important input variables for the NN output of the final $\btag$ meson include: the network outputs of the daughter particles; the reconstructed masses of the daughters; ${\Delta E = E_{\btag} - \ebeam}$, which is the difference between the $\btag$ candidate energy and the beam energy in the center-of-mass system (CMS); the mass difference between $M(D^*)$ and $M(D)$; the angles between the daughters in the $\btag$ meson rest frame; the momentum of the daughters in the lab frame; and $\cos\Theta_B$, the cosine of the angle between the beam and the $\btag$ direction. The network output can be interpreted as the probability that the $\btag$ candidate is correctly reconstructed, which means all particle hypotheses of the decay chain are correct. In the case of multiple $\btag$ candidates, the candidate with the highest network output is selected. 
\par
For the network output, differences between data and MC have been observed~\cite{sibidanov13}; $\btag$ decay modes with at least two pions in the final state show the largest deviation. In charmed semileptonic signal-side $B$ decays the efficiency in MC is overestimated by approximately one third. From that, a correction factor depending on the hadronic tag-side decay channel is obtained, and it is applied to all MC samples used in the analysis.
\par 
For the analysis, additional event shape variables are added to the network training. The variables are used to discriminate between spherical $\bbar$ and jet-like $\qq$ continuum processes. The event shape variables are modified Fox-Wolfram moments~\cite{fox78} and the thrust axis of the $\btag$ meson candidate.

\section{Selection}
\subsection{Missing mass}
With the $\btag$ candidate three-momentum $\vec{\,p}_{\btag}$, the four-momentum of the signal-side $\bsig$ meson  in the CMS is given by $p_{\bsig} = (\ebeam/c, -\vec{\,p}_{\btag})$. This makes use of the two-body decay kinematics of the $\Upsilon(4\textrm{S})$ and the measured CMS boost of the $\bbar$ system. The $\bsig$ four-momentum is used to compute the squared missing mass, which is the strongest discriminator between signal and background. The variable is defined as
\begin{equation*}
 \miss=(p_{\bsig}-p_{\ell}-p_{\gamma})^{2}/c^4,
\end{equation*}
where the four-momenta of the daughter lepton and photon are subtracted from that of the $\bsig$ candidate. For correctly reconstructed signal events, the variable corresponds to the neutrino mass and therefore peaks around zero. The resolution of this signal peak is improved by using $\ebeam$ instead of $E_{\btag}$ in $p_{\bsig}$. An additional improvement in resolution is achieved for $\Benu$ decays by taking bremsstrahlung into account: the four-momentum of the signal electron candidate is corrected by the addition up to photon below an energy of 1\,GeV within a five degree cone around the direction of its momentum. For the signal extraction, the region with ${\miss \in (-2.0, 4.0)\, \textrm{GeV}^2/c^4}$ around the signal peak is used.
\par
The analysis begins with a selection with high signal efficiency and purity, followed by a signal-yield extraction with a fit to the missing mass in bins of a NN output. The number of network-output bins as well as the selection of variables used in the training of the network are optimized for signal significance. With the exception of the lepton identification (ID), the selection is identical for both $\Benu$ and $\Bmunu$. 
\subsection{Tag-side selection}
For the $\btag$ candidate, the beam-energy-constrained mass ${\mbc = \sqrt{\ebeam^{\,2} - \vec{\,p}_{\btag}^{\,2}}}\,\,/c^2$ is required to be greater than 5.27\,GeV/$c^2$. A selection of ${\Delta E \in (-0.15, 0.10)\,\textrm{GeV}}$ is applied; this variable is not used elsewhere since it is strongly correlated with the missing mass. A loose selection on the network output of the fully reconstructed $\btag$ meson is chosen to have a probability above $2 \times 10^{-4}$ of being correctly reconstructed.

\begin{figure*}[!htb]
   \subfloat[Electron channel]{\includegraphics[scale=0.7]{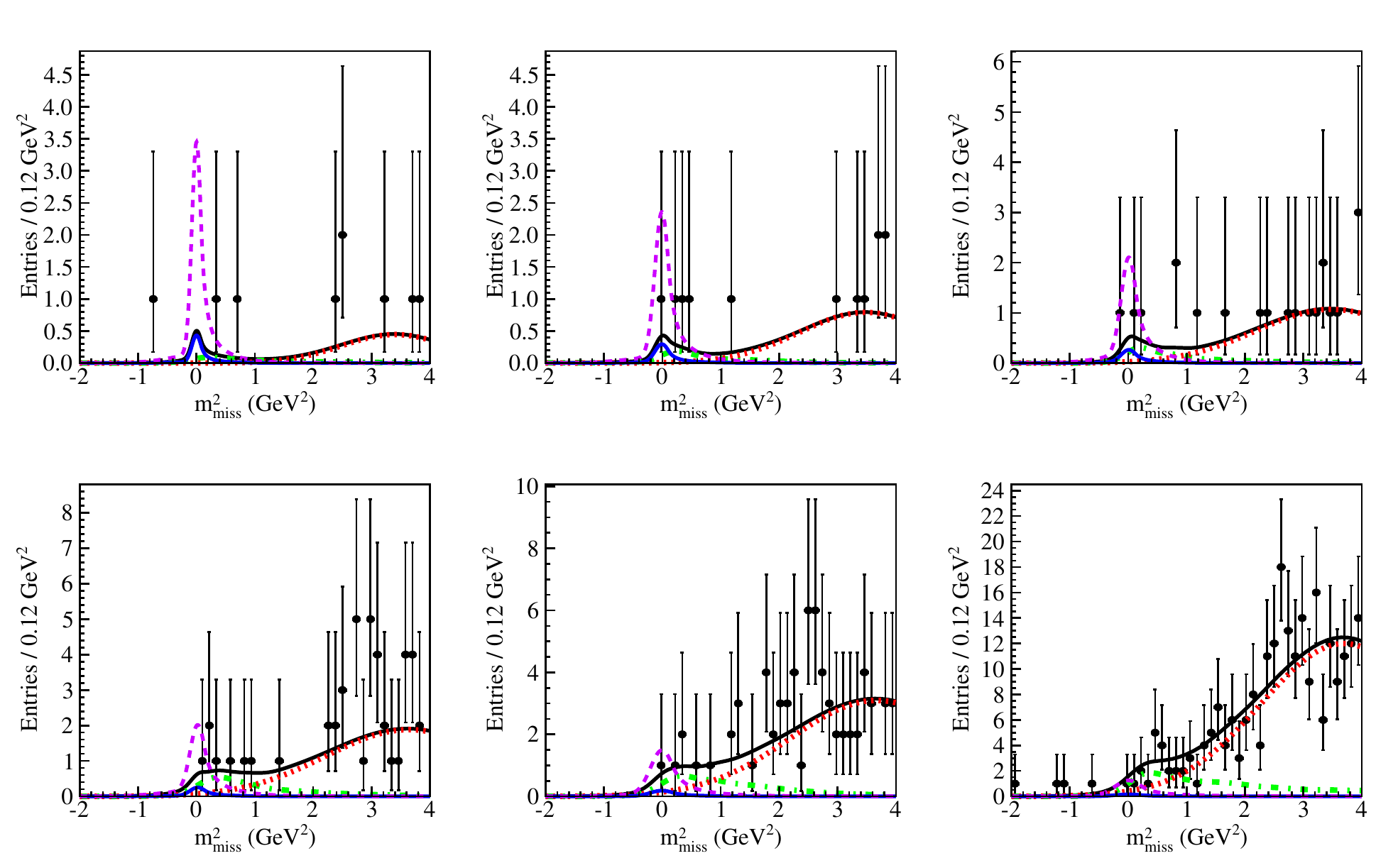}}    \\  
   \subfloat[Muon channel]{\includegraphics[scale=0.7]{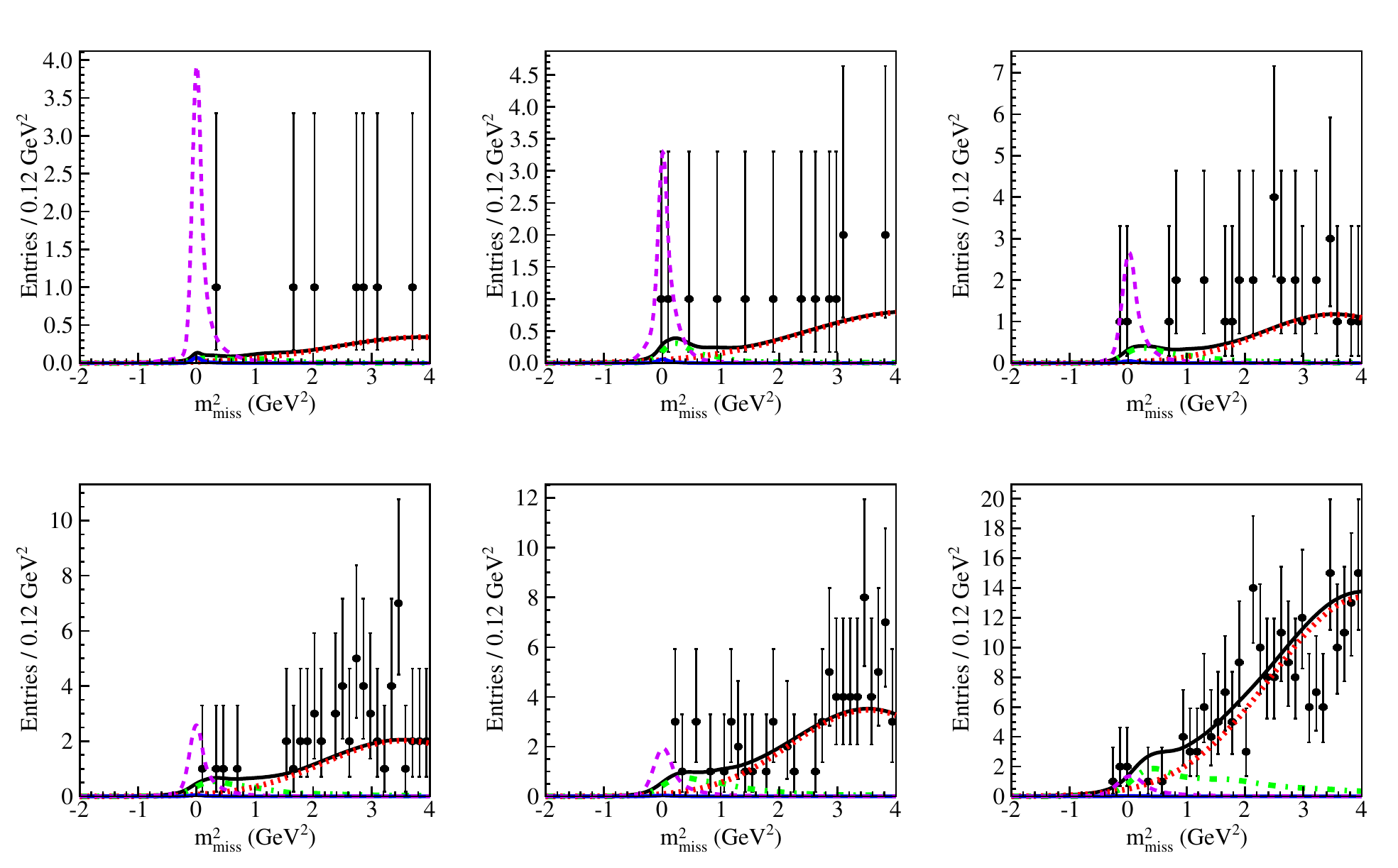}}         
  \caption{(color online) Distributions of $\miss$ on data (points with error bars) in bins of the network output. The PDFs are for signal (solid blue), enhanced signal (dashed violet), fixed $\BXlnu$ backgrounds (dash-dotted green), fitted backgrounds (dotted red), and the sum (solid black). The enhanced signal function, which has the same normalization for each bin, corresponds to a branching fraction of $30\times10^{-6}$. The most signal-like bin is found in the upper left panel. Proceeding from left to right, the distributions become increasingly more background-like and the most background-like bin is shown in the lower right panel.}
  \label{fig:binnedMeasurement}
\end{figure*}

\begin{figure}[!htb]
 \subfloat[Electron channel]{\includegraphics[width=.5\textwidth]{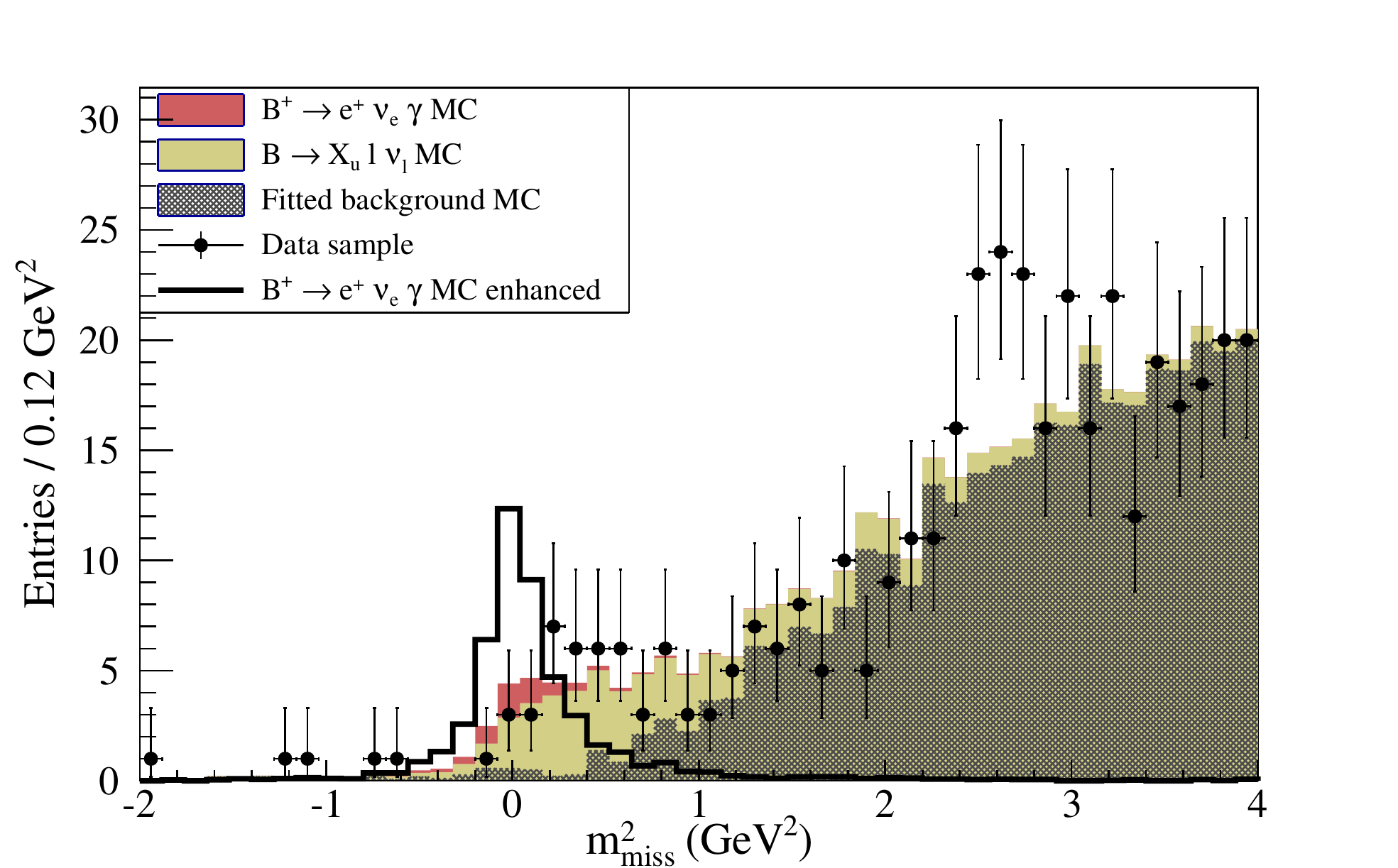}}\\
 \subfloat[Muon channel]{\includegraphics[width=.5\textwidth]{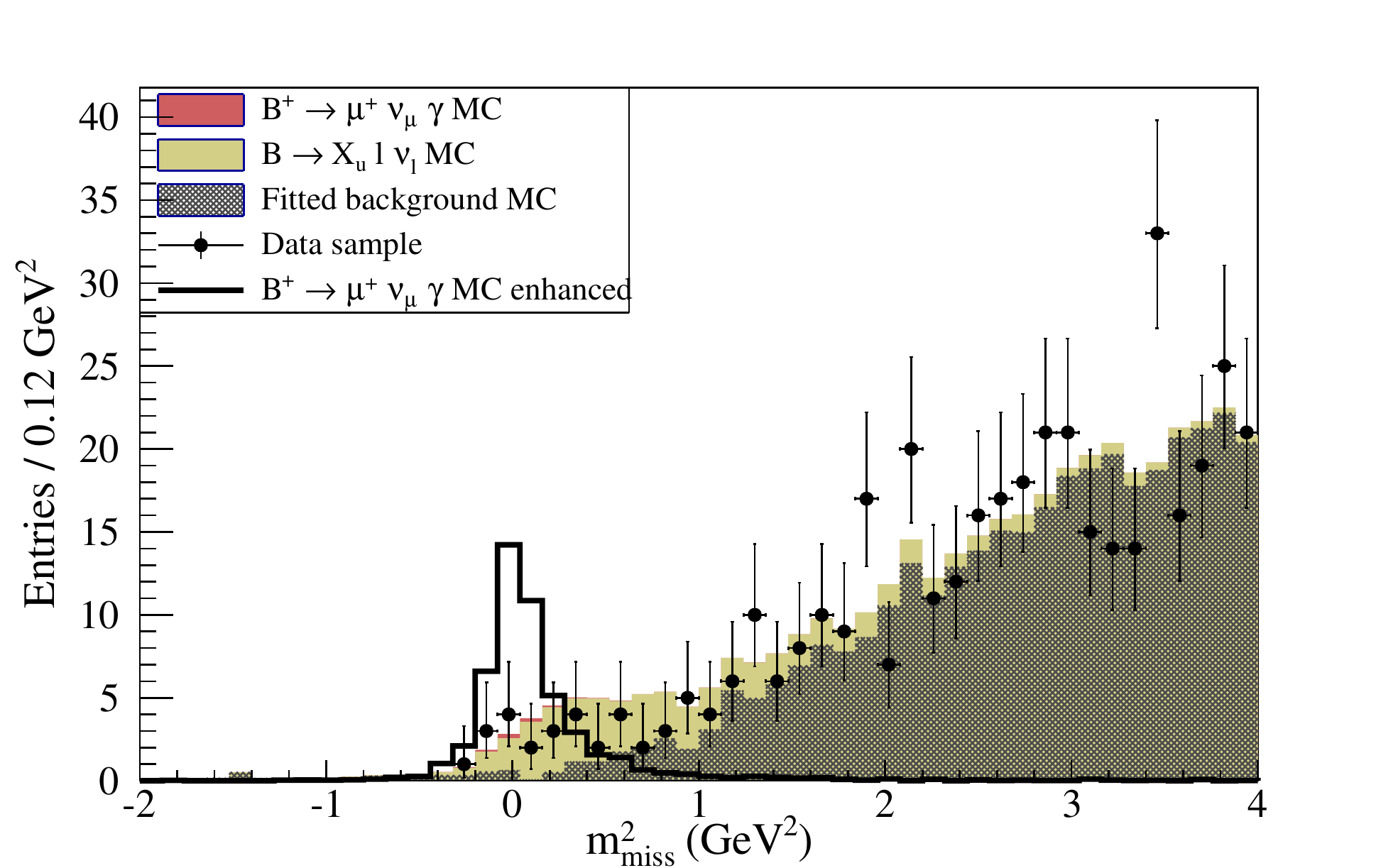}}
 \caption{(color online) Unbinned $\miss$ distribution where the enhanced signal corresponds to a branching fraction of $30\times10^{-6}$.}
\label{fig:unbinnedMeasurement}
\end{figure}

\begin{figure}[!htb]
 \subfloat[Electron channel]{\includegraphics[width=.5\textwidth]{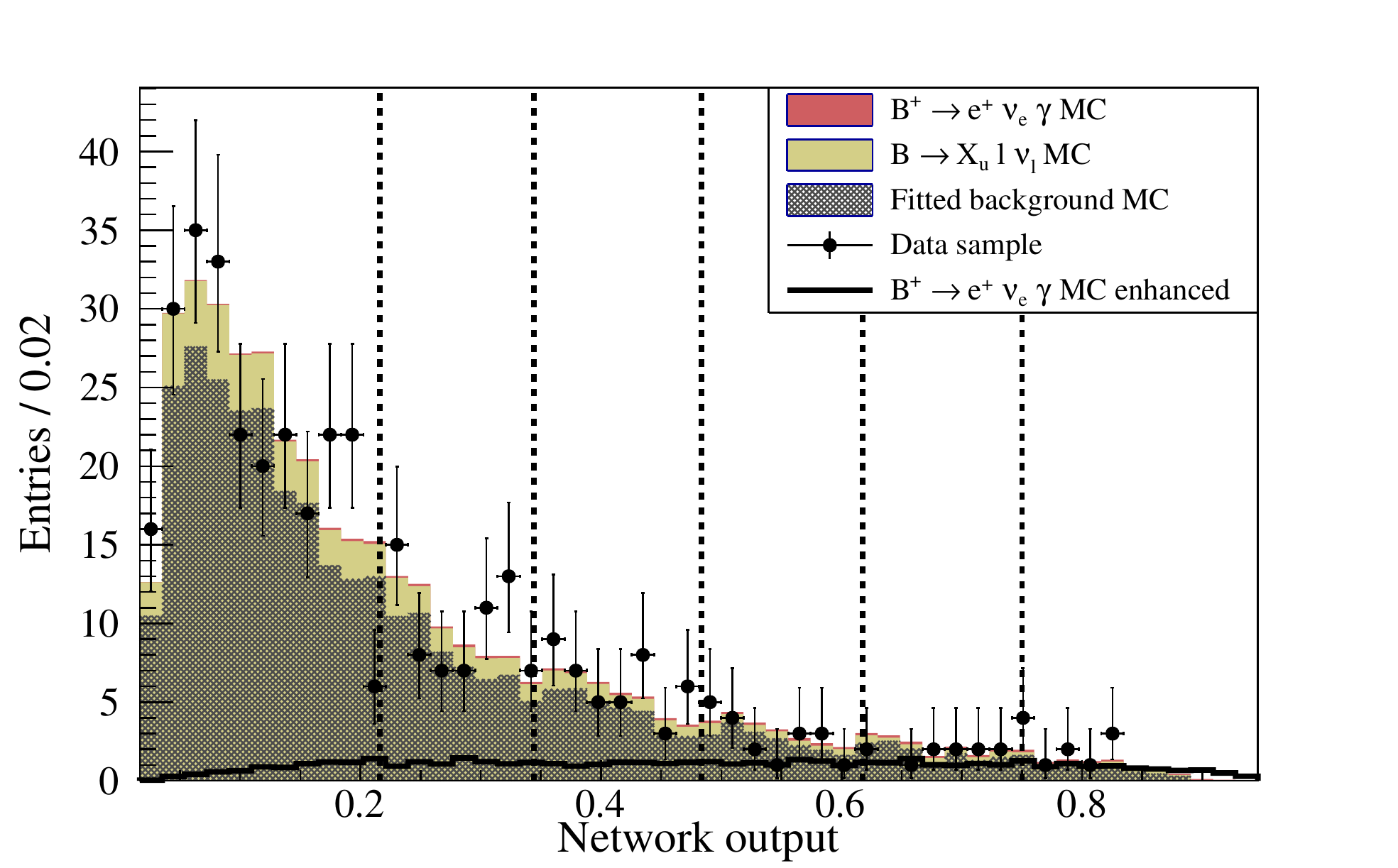}}\\
 \subfloat[Muon channel]{\includegraphics[width=.5\textwidth]{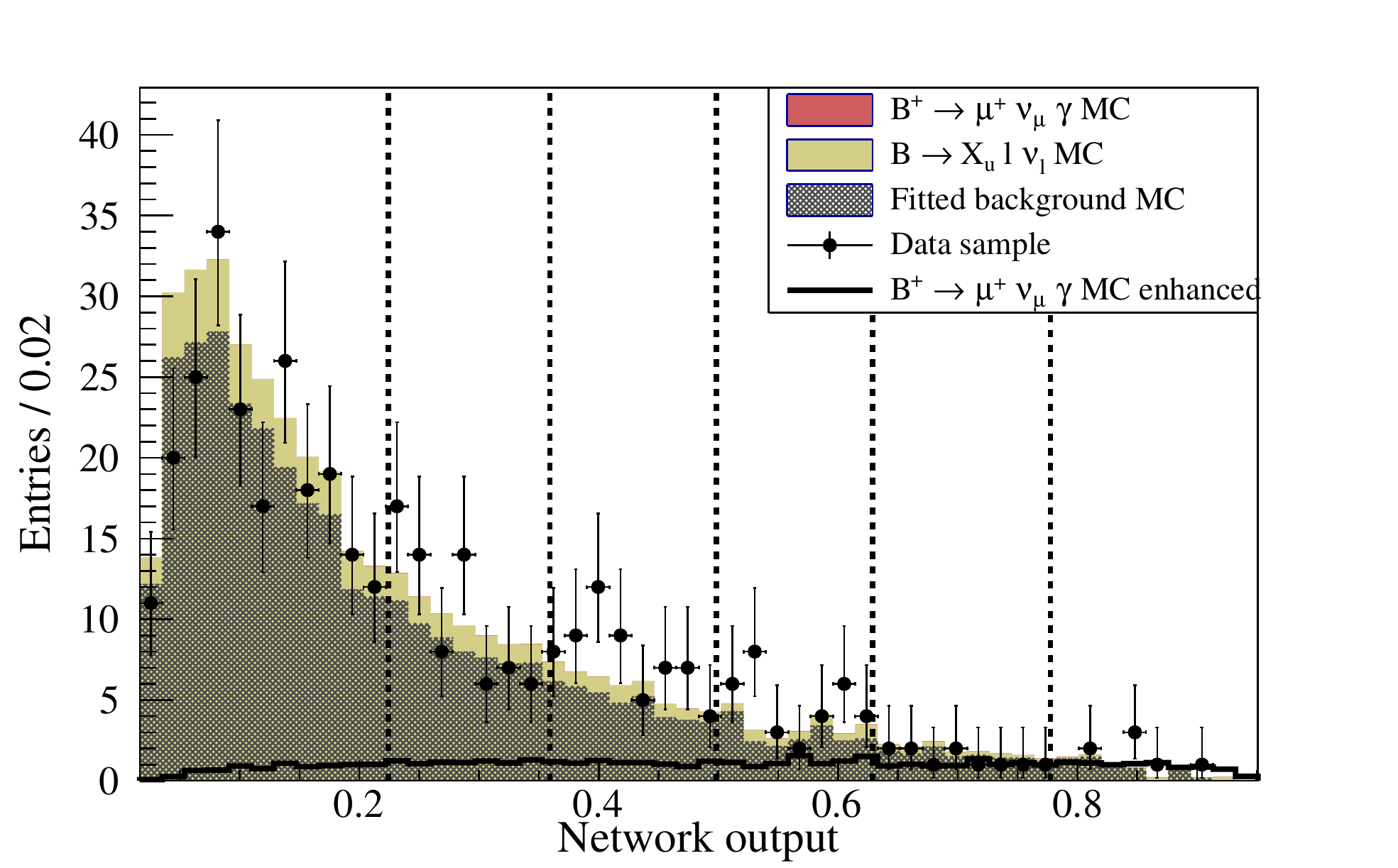}}
 \caption{(color online) Network outputs used for $\miss$ binning where the bin boundaries are indicated by the dashed lines. The normalizations of the MC distributions are taken from the fit results in $\miss$ and the enhanced signal corresponds to a branching fraction of $30\times10^{-6}$.}
\label{fig:netout}
\end{figure}

\subsection{Signal-side selection}
After hadronic tag-side reconstruction, one charged track and one high-energy photon are expected in the detector. No additional charged tracks beyond the signal's lepton daughter are permitted. The signal-side charged-track selection demands the same selection for the impact parameters as the tag-side: ${dr < 2\,\textrm{cm}}$ and ${|dz| < 4\,\textrm{cm}}$. The charge of the signal lepton candidate is required to be opposite that of the $\btag$. Curling tracks, which can be counted twice, are taken into account on the signal side by counting two tracks as one if the cosine of the angle between them is above 0.999 and their transverse momentum differs by less than 30\,MeV/$c$.
\par 
Electrons are identified from a likelihood formed with information from multiple detectors: the energy loss in the CDC; the ratio of energy deposition in the ECL to the track momentum; the shower shape in the ECL; the matching of the charged track to the shower position in the ECL; and the photon yield in the ACC~\cite{eid}. Muons are identified from charged tracks extrapolated to the outer detector; the difference between the expected and measured penetration depth of the track as well as the transverse deviation of KLM hits from the extrapolated track are used to distinguish muons from hadrons~\cite{muid}. Adding the particle ID to the final selection, 95\% (99\%) of events with a wrong-lepton hypothesis are vetoed with a reduction in signal selection efficiency of about 2\% (1.2\%) for the muon (electron) channel.
\par
The analysis is performed with two energy thresholds of 1\,GeV and 400\,MeV for the signal photon candidate in the $\bsig$ rest frame, where the most energetic photon in the $\bsig$ rest frame is identified as the signal photon candidate. The 1\,GeV threshold is a lower bound for which the theoretical model is valid; however, a secondary analysis covering a larger phase space is performed, with a 400\,MeV bound chosen to remove the divergent part in the decay model at lower energies. The missing momentum in the event $\pnusig$ has to be above 800\,MeV/$c$ in the $\bsig$ rest frame, to be consistent with the presence of a high energy neutrino.
\par
Events in which a signal photon candidate is mis-reconstructed from bremsstrahlung radiation originating from the signal electron are vetoed by requiring that the cosine of the angle between the lepton and photon candidates in the $\bsig$ rest frame ($\coslg$) lie below 0.6. For the cosine of the angle between the missing momentum and the signal photon candidate in the $\bsig$ rest frame ($\cosgnu$) a discrepancy is observed between MC and data for values below $-0.9$ in the sideband of ${\mbc < 5.27\, \textrm{GeV}/c^2}$; therefore $\cosgnu$ is selected to be larger than $-0.9$. The remaining energy in the ECL is the summed energy of clusters not associated with signal or tag-side particles and is required to be below 900\,MeV. Here, clusters are required to have energies above of 50, 100, and 150\,MeV for the barrel, forward, and backward end-cap calorimeter, respectively. These energy thresholds with directional dependence are proven to veto background in the detector not related to physical processes. 
\par
To suppress the main background of $\Bpilnu$ decays, a $\pi^0$ veto is constructed that combines the signal photon candidate with all remaining photons in the ECL above an energy of 100\,MeV to compute the invariant mass, where only the candidate closest to the nominal $\pi^0$ mass is kept. A $\pi^0$ mass is only computed if at least one remaining photon above an energy of 100\,MeV is left in the ECL. The number of events with a computed $\pi^0$ mass decreases with a rising energy threshold, as does the number of events vetoed by a selection on the resulting mass spectrum. On the other hand, an increasing energy threshold improves the signal and background separation since fewer photons are combined with the signal photon candidate. This reduces the possibility of calculating a $\pi^0$ mass close to the correct one by chance. The 100\,MeV threshold is chosen to ensure a high signal efficiency of about 99\% while achieving a good background rejection of 45\% for $\Bpilnu$ processes, when a window of 30\,MeV/$c^2$ around the nominal $\pi^0$ mass is vetoed.
\par
The overall signal selection efficiency after full reconstruction is 47\% (45\%) for the muon (electron) channel. The expected event numbers from the background MC samples are: 328 (299) for $b \to c$ decays, 78 (76) for $\bulnu$ decays, and 17 (6) events from non-resonant $\qq \to (u, d, s, c)$ processes for the muon (electron) channel. The contribution from $b \to s$ processes is found to be negligible. 

\subsection{Neural network training}
To further optimize the signal selection, another NN is formed with the NeuroBayes package~\cite{feindt06}. This software computes each input variable's significance from the training; this is used to retain only the most significant variables in the network. The variables included in the training are: the extra energy in the ECL, $\coslg$, and $\cosgnu$. To further separate the main background processes of $\Bpilnu$ and $\Betalnu$, where the $\pi^0$ and $\eta$ decay into two photons and one of the photons is misidentified as the signal photon, meson-veto variables are incorporated into the network. These are computed in the same way as for the selection above but with different energy thresholds on the remaining photons in the ECL. 
\par
The thresholds are increased in 10\,MeV steps from 20 to 100\,MeV. The number of photons combined with the signal photon candidate depends on this energy threshold, and since only the combination closest to the nominal mass is taken into account, different photon combinations end up in the mass spectrum. This leads to different invariant mass spectra with complementary information. The $\eta$ invariant mass is computed in the same way, with energy thresholds between 20 and 300\,MeV. Only the six most significant meson masses are retained in the training.
\par
Signal MC samples of both signal channels are trained simultaneously against the $\bulnu$ MC and the high-luminosity $\Bpilnu$ MC sample. For the secondary analysis with $\gsig\,>$\,400\,MeV, the angles $\coslg$ and $\cosgnu$ are excluded from the training to reduce the signal model dependence of the result.  
\section{Signal extraction}
\subsection{Fit model}
The signal yield is determined by an extended unbinned maximum likelihood fit to the $\miss$ distribution in six bins of the NN output. The likelihood function is given by 
\begin{equation*}
\textrm{ln}\,{\cal{L}} = \sum_{j=1}^{N_\textrm{tot}}\textrm{ln}\Big\{\sum_i^{N_c} N_i{\cal{P}}_i(\miss, n_\textrm{out}) \Big\} -\sum_i^{N_c} N_i,
\end{equation*}
where $N_\textrm{tot}$ is the total number of events in the data set, $N_c$ denotes the number of components in the fit, $N_i$ is the number of events for the $i^\textrm{th}$ component, and ${\cal{P}}_i$ represents the probability density function (PDF) for that component as a function of $\miss$ and the network output $n_\textrm{out}$. 
\par
The fit model consists of three components: ${\Blnu}$ signal; measured ${\bulnu}$, decays referred to hereinafter as the ${\BXlnu}$ component; and a component denoted as ``fitted background'' that includes unmeasured ${\bulnu}$ contributions, resonant ${b \to c}$ decays, and non-resonant $\qq$ processes. In the fit to data, the expected yield of the ${\BXlnu}$ component containing the known decay modes with $X_{u} = \pi^{0}$, $\eta$, $\omega$, $\rho^{0}$, $\pi^{+}$, $\rho^{+}$, and $\eta'$ is fixed according to the world average values of the branching fractions~\cite{olive14}. The shapes of the three components are determined from MC in each network output bin separately and fixed in the fit to data together with the relative normalizations among the bins. The PDF for the $i^\textrm{th}$ component is given by
\begin{equation*}
{\cal{P}}_i(\miss, n_\textrm{out}) = f^{n_\textrm{out}}_i{\cal{P}}^{n_\textrm{out}}_i(\miss),
\end{equation*}
where $f^{n_\textrm{out}}_i$ denotes the fixed fraction of $N_i$ events in the bin and ${\cal{P}}^{n_\textrm{out}}_i$ is the PDF in that NN bin with central value $n_{\textrm{out}}$. 
\par
By design, each bin contains the same number of expected signal events and the bin boundaries are shown in Fig.~\ref{fig:netout}. The number of network output bins is chosen to maximize the expected significance of the signal, which is determined in toy MC studies. The number of signal and fitted background events are the two free parameters of the fit model. The two signal channels $\Benu$ and $\Bmunu$ are measured in separate fits. A simultaneous fit to both channels is performed to measure the $\Blnu$ branching fraction. Lepton universality is assumed for the latter measurement, where the signal branching fractions of the two channels are fixed to the same value. To avoid a fit bias, all yields are unconstrained and negative values are allowed in the fit.
\par 
The signal component is parametrized with the sum of a Crystal Ball function~\cite{skwarnicki86} and a Gaussian with a common mean. A shape for the fitted background component is given by an exponential with a polynomial in its argument
\begin{equation*}
f(x; x_{0}, \alpha, \beta) =  e^{\alpha(x-x_{0})^{2} + \beta (x-x_{0})}.
\end{equation*}
The fixed background component of ${\BXlnu}$ decays is modeled with a non-parametric PDF using a kernel estimation algorithm~\cite{cranmer00}, where each data point is represented by a Gaussian and their sum yields a probability density function. The width of the Gaussian kernels is a parameter of the algorithm that is chosen to produce a smooth description of the MC. Identical functions are fitted for both signal channels.
\subsection{Significance and limit determination}
The significance of the signal is defined as ${\sqrt{-2\textrm{ln}({\cal{L}}_b/{\cal{L}}_{(s + b)})}}$ where ${{\cal{L}}_b}$ and ${{\cal{L}}_{(s + b)}}$ are the maximum likelihood value of the background and signal plus background model, respectively. The maximum likelihood values for null and signal hypothesis are obtained from the likelihood profile, where both likelihood values are taken from the same data distribution. An upper limit at 90\% credibility level~\cite{BayesianStat} is determined from an integration of the likelihood function up to the 90\% quantile, where only the range for positive signal yields is used. The systematic uncertainty is included by convolving the likelihood function with a Gaussian whose width is equal to the systematic error. Systematic errors affecting only the signal yield are included in the determination of the significance. The total systematic error, including errors impacting the overall yield, is used for the measurement of the branching fraction and its upper limit. Since the systematic errors are asymmetric, the downward errors are used for the significance and the upward errors for the upper limit. The expected fit results from an average over many toy MC studies are listed in Table~\ref{tab:measurement} for the nominal and secondary analyses. The expected signal yield depends on the value of $\lb$. The expected fit significances are determined with a signal branching fraction of ${5 \times 10^{-6}}$ and the expected upper limits are measured without any signal contribution. For the simultaneous fit, a significance of 2.9$\sigma$ including systematic errors is expected.

\begin{table*}[!ht]
{\renewcommand{\arraystretch}{1.5}
\begin{tabular}{| c | c | c | c " c | c | c | c |}
 \cline{2-8}
 \multicolumn{1}{c|}{} & \multicolumn{7}{c|}{Nominal analysis with $\gsig\,>$\,1\,GeV} \\\cline{2-8}
 \multicolumn{1}{c|}{} & \multicolumn{3}{c"}{MC expectation}  & \multicolumn{4}{c|}{Data measurement} \\\hline
  Mode & Yield & Significance $(\sigma)$ & ${\cal{B}}$ limit $(10^{-6})$ & Yield & ${\cal{B}}\,\,(10^{-6})$ & Significance $(\sigma)$ & ${\cal{B}}$ limit $(10^{-6})$ \\
\hline
$\Benu$ & 8.0 $\pm$ 4.5 $^{+1.0}_{-1.3}$ & 2.1 & $<7.5$ & $6.1^{+4.9\,+1.0}_{-3.9\,-1.3}$ & $3.8^{+3.0\,+0.7}_{-2.4\,-0.9}$ & 1.7 & $<\,6.1$\\
$\,\,\Bmunu$ & 8.7 $\pm$ 4.6 $^{+1.0}_{-1.5}$ & 2.2 & $<6.9$ & $0.9^{+3.6\,+1.0}_{-2.6\,-1.5}$ & $0.6^{+2.1\,+0.7}_{-1.5\,-1.1}$ & 0.4 & $<\,3.4$\\
$\Blnu$ & 16.5 $\pm$ 6.5 $^{+1.6}_{-2.2}$ & 2.9 & $<4.8$ & $6.6^{+5.7\,+1.6}_{-4.7\,-2.2}$ & $2.0^{+1.7\,+0.6}_{-1.4\,-0.7}$ & 1.4 & $<\,3.5$\\
  \hline 
 \end{tabular}
\begin{tabular}{| c | c | c | c " c | c | c | c |}
  \multicolumn{8}{c}{} \\
  \multicolumn{8}{c}{} \\
  \cline{2-8}
 \multicolumn{1}{c|}{} & \multicolumn{7}{c|}{Secondary analysis with $\gsig\,>$\,400\,MeV} \\\cline{2-8}
 \multicolumn{1}{c|}{} & \multicolumn{3}{c"}{MC expectation}  & \multicolumn{4}{c|}{Data measurement} \\\hline

  Mode & Yield & Significance $(\sigma)$ & ${\cal{B}}$ limit $(10^{-6})$ & Yield & ${\cal{B}}\,\,(10^{-6})$ & Significance $(\sigma)$ & ${\cal{B}}$ limit $(10^{-6})$ \\
\hline
$\Benu$ & 12.4 $\pm$ 6.2 $^{+1.8}_{-2.3}$ &  2.1 & $<6.8$ & $11.9^{+7.0\,+1.8}_{-6.0\,-2.3}$ & $4.9^{+2.9\,+0.8}_{-2.5\,-1.0}$ & 2.0 & $<\,9.3$\\
$\,\,\Bmunu$ & 11.9 $\pm$ 6.0 $^{+1.7}_{-2.1}$ & 2.2 & $<6.2$ & $-0.1^{+5.2\,+1.7}_{-4.1\,-2.1}$ & - & - & $<\,4.3$ \\
$\Blnu$ & 24.9 $\pm$ 8.7 $^{+3.0}_{-3.5}$ & 2.9 & $<4.3$ & $11.3^{+8.4\,+3.0}_{-7.4\,-3.5}$ & $2.3^{+1.7\,+0.7}_{-1.5\,-0.8}$ & 1.4 & $<\,5.1$ \\
  \hline
   \end{tabular}
   }
   \caption{Expected signal yields obtained from MC for ${\cal B}(\Blnu) = 5 \times 10^{-6}$ and measured signal yields on data, where the first error is statistical and the second error systematic. The significances and credibility levels contain systematic errors. The credibility levels are given at 90\% where the expected MC limit is determined without signal.}
\label{tab:measurement}
\end{table*}

\begin{table*}
{\renewcommand{\arraystretch}{1.3}
\begin{tabular}{| c | c | c " c | c |}
  \cline{2-5}
  \multicolumn{1}{c|}{} & \multicolumn{2}{c"}{Nominal analysis with $\gsig\,>$\,1\,GeV} & \multicolumn{2}{c|}{Secondary analysis with $\gsig\,>$\,400\,MeV}  \\
  \hline
  Mode & MC expectation & Measured yield & MC expectation & Measured yield \\
  \hline
$\Benu$ & $315 \pm 4.2$ & $336^{+20}_{-19}$ & $668 \pm 6.1$ & $739^{+29}_{-28}$ \\
$\,\,\Bmunu$ & $348 \pm 4.5$ & $352^{+20}_{-19}$ & $714 \pm 6.4$ & $759^{+29}_{-28}$ \\
  \hline 
  \end{tabular}
  }
   \caption{Fitted background yields compared to the MC prediction with statistical errors only.}
\label{tab:measurementBkg}
\end{table*}

\begin{table}[!ht]
{\renewcommand{\arraystretch}{1.3}
\begin{tabular}{| l | c | c |}
  \hline
  Source & $\Bmunu$ &  $\Benu$ \\
  \hline
  \hline
  Fit shapes & $^{+0.75}_{-1.34}$ & $^{+0.64}_{-1.06}$ \\
  Meson veto network & $\pm0.58$ & $\pm0.66$ \\
  Fixed $\BXlnu$ yield & $\pm0.18$ & $\pm0.24$ \\
  $\Blnu$ model & $-0.01$ & $-0.05$ \\
  \hline
  Additive Error & $^{+0.97}_{-1.47}$ & $^{+0.95}_{-1.27}$ \\
  \hline
  \hline
  Lepton ID & $\pm0.42$ & $\pm0.18$ \\
  Tag-side efficiency & $\pm0.35$ & $\pm0.34$ \\
  Tag-side NN & $\pm0.13$ & $\pm0.40$ \\
  Tracking efficiency & $-0.01$ & $-0.01$ \\
  N$_{\bbar}$ & $\pm0.11$ & $\pm0.11$ \\
  \hline
  Multiplicative Error & $\pm0.57$ & $\pm0.55$ \\
  \hline
  \hline
  Combined Error & $^{+1.12}_{-1.58}$ & $^{+1.10}_{-1.39}$ \\
  \hline
  \multicolumn{3}{c}{} \\
  \hline
  \multicolumn{1}{|l|}{Source} & \multicolumn{2}{c|}{$\Blnu$} \\
  \hline
  \hline
  Additive Error & \multicolumn{2}{c|}{$^{+1.64}_{-2.15}$} \\
  Multiplicative Error & \multicolumn{2}{c|}{$\pm0.99$} \\
  Combined Error & \multicolumn{2}{c|}{$^{+1.92}_{-2.37}$} \\
  \hline
\end{tabular}
}
\caption{Systematic uncertainties on the signal yield grouped by error-types for the nominal analysis with ${\gsig > 1\,\textrm{GeV}}$. Deviations are given in signal yields.}
\label{tab:sysErr} 
\end{table}
\subsection{Toy MC and sideband data checks}
The fit model is checked for a bias in extended toy MC studies where, pull distributions are used to quantify the size of the bias. The pull distributions are computed from the deviation from the true value divided by the fit error and have a standard normal distribution for unbiased fits. This is used in a linearity test of the signal yield, which checks whether the bias of the fit results depends on the signal branching fractions. The pull distributions are in agreement with standard normal distributions, indicating no bias for branching fractions that result in a significant measurement. A test of the credible interval~\cite{BayesianStat} coverage counts the number of events for which the true value is contained inside the 90\% interval. For a branching fraction of ${5\times10^{-6}}$, 95\% of the true values are contained inside the interval; this number increases to more than 99\% below a branching fraction of ${3\times10^{-6}}$. Since the likelihood is only integrated for positive signal yields to determine the limit, the 90\% quantile is moved to higher values. Therefore, the upper limit is a conservative measure. The same results are found for the secondary analysis. 
\par
The background MC shapes are compared to data in the ${\mbc < 5.27}$\,GeV/$c^2$ sideband. Additionally, the agreement of the input variables to the NN is checked in a ${\BXlnu}$ enhanced region of ${\miss \in (0.3, 1.0)\,\textrm{GeV}^2/c^4}$ and a generic background dominated region of ${\miss \in (1.0, 4.0)\,\textrm{GeV}^2/c^4}$. All considered distributions agree between data and MC, except for the previously mentioned discrepancy in the $\cosgnu$ distribution. 

\section{Measurement}
The fit results are listed in Table~\ref{tab:measurement} and the $\miss$ distributions for the nominal analysis are shown in Fig.~\ref{fig:binnedMeasurement} for both signal channels. No significant signal is found in any of the fits. To offer a better overview of the fit results, unbinned distributions of the results are shown in Fig.~\ref{fig:unbinnedMeasurement}. Good agreement between data and MC for the network output is shown in Fig.~\ref{fig:netout}. The fitted background yields in the data are in agreement with the MC prediction, as shown in Table~\ref{tab:measurementBkg}. Assuming that only a few signal events are found below the photon energy threshold of 400\,MeV, the partial branching fractions of the secondary analysis can be compared to the BaBar measurement~\cite{aubert09} for the whole energy range. Limits on $\lb$ are computed by integrating the differential decay width from Equation~\ref{eq:singlediff}
\begin{equation*}
 \Delta {\cal{B}} = \frac{\tau_{B_d}}{\hbar} \int\limits_{1 \textrm{GeV}}^{m_B/2c^2} dE_{\gamma} \frac{d\Gamma}{dE_{\gamma}}
\end{equation*}
and solving for $\lb$, where the integral includes the partial phase space ${\gsig > 1\,\textrm{GeV}}$ up to half of the $B$ meson mass. The input parameters for the differential decay width are taken from Ref.~\cite{beneke11} and the value for the soft correction $\xi(E_{\gamma})$ is taken from Ref.~\cite{braun13}. All parameters are varied by their uncertainties to obtain parameter combinations yielding minimal and maximal values for $\lb$. With the $\Blnu$ limit of the nominal analysis, a central value ${\lb > \lambdalimit\,\textrm{MeV}}$ is obtained at 90\% credibility level. The limit changes within a range of ${\lb > (\lambdallimit, \lambdaulimit) \textrm{MeV}}$ with varying input parameters.\footnote{Several values of $\xi(E_{\gamma})$ are calculated in Ref.~\cite{braun13} for different true values of $\lb$. We identify the central value of $\xi(E_{\gamma})$ with the one obtained for ${\lb = 300\,\textrm{MeV}}$. To obtain the error on $\xi(E_{\gamma})$, the whole range of true values for $\lb$ is taken into account.} Similar values are obtained for the secondary analysis.

\section{Systematic uncertainties}
Systematic errors are estimated in toy MC studies where the default and the varied fit models are applied to the same toy sample and the difference in signal yield is taken as a systematic deviation averaged over many toy measurements. The results are shown in Table~\ref{tab:sysErr} for the nominal analysis.
\par
The largest error is given by the variation of the fit shapes, where the 1$\sigma$ fit error from MC is varied. For the non-analytical shape obtained from the kernel estimator algorithm, the size of the Gaussian kernels is varied to obtain a considerable shape variation. 
\par
The systematic error on the meson-veto network is obtained from the control channel $B^0\to K^{*0}\gamma$. Here, the signal photon candidate is combined with the remaining photon candidates to compute the meson mass spectra and obtain the network output distribution. From this distribution, a double ratio of data and MC is calculated as ${(N_i^\textrm{MC}/N_\textrm{sum}^\textrm{MC})/(N_i^\textrm{data}/N_\textrm{sum}^\textrm{data})}$, where $N_i$ is the event count in the $i^\textrm{th}$ bin and $N_\textrm{sum}$ the total number of events. The largest deviation between data and MC is found to be 8\% in the most background-like network output bin. An alternate model is obtained by using the double ratio values to reweight the binned $\miss$ distribution in $\Blnu$. The angles $\coslg$ and $\cosgnu$, as well as the remaining energy in the ECL, cannot be used in the NN trained on the control sample. Therefore, a separate network without these variables is trained on the $\Blnu$ samples, which is then used to obtain the double ratios in the control channel. 
\par
The fixed yields of the measured $\BXlnu$ backgrounds are varied by their world-average errors~\cite{olive14}. The systematic uncertainty related to the $\Blnu$ decay signal model is estimated by comparing the latest NLO model~\cite{beneke11} with an older LO calculation~\cite{korchemsky00}. Here, the shape difference in the $\miss$ distribution is found to be small and parametric errors of the theory are also found to have a negligible effect on the branching fraction determination. 
\par
The systematic uncertainty related to lepton ID is determined in $\gamma\gamma\to\ell^+\ell^-$ processes and the error is found to be 2.2\% and 5.0\% for electrons and muons, respectively. The error for the tag-side efficiency has been determined in Ref.~\cite{sibidanov13} to be 4.2\%. The error for the tag-side NN is taken from the sideband ${\miss > 0.3\,\textrm{GeV}^2/c^4}$, where the difference in the data-MC selection efficiency is taken as a systematic error. Systematic deviations for the tracking efficiency are determined with high transverse momentum tracks from partially reconstructed $D^*$ mesons; the deviation is $-0.13\%$. 
\par
To obtain the systematic error for the simultaneous fit to both channels, all errors are assumed to be fully correlated except for the errors on the fit shapes and the lepton ID, for which no correlation is assumed. The total systematic error is less than half of the statistical error.

\section{Conclusion}
In summary, we report the upper limits of the partial branching fraction with ${\gsig > 1\,\textrm{GeV}}$ for semileptonic $\Blnu$ decays with the full Belle data set of ${(771.6 \pm 10.6) \times 10^{6} \bbar}$ pairs. The signal photon energy requirement ensures a reliable theoretical description of the decay process. The results at 90\% credibility level are
\begin{eqnarray*}
    & {\cal B}(\Benu) & <  \BRelimit \times 10^{-6},  \\
    & \,{\cal B}(\Bmunu) & < \BRmulimit \times 10^{-6},  \\ 
    & {\cal B}(\Blnu) & < \BRllimit \times 10^{-6}.  
\end{eqnarray*}
These results improve the limits measured by BaBar~\cite{aubert09}. The limit of the combined channel $\Blnu$ translates into a boundary of ${\lb > \lambdalimit\,\textrm{MeV}}$ at 90\% credibility level, where this limit evolves within the range ${\lb > (\lambdallimit, \lambdaulimit)\,\textrm{MeV}}$ by varying the input parameters of the decay width. A secondary analysis with a lower signal photon energy threshold of ${\gsig > 400 \textrm{MeV}}$ gives consistent results.

%
\section*{Acknowledgements}
We thank the KEKB group for the excellent operation of the accelerator; the KEK cryogenics group for the efficient operation of the solenoid; and the KEK computer group, the National Institute of Informatics, and the PNNL/EMSL computing group for valuable computing and SINET4 network support.  We acknowledge support from the Ministry of Education, Culture, Sports, Science, and Technology (MEXT) of Japan, the Japan Society for the Promotion of Science (JSPS), and the Tau-Lepton Physics Research Center of Nagoya University; the Australian Research Council and the Australian Department of Industry, Innovation, Science and Research; Austrian Science Fund under Grant No.~P 22742-N16 and P 26794-N20; the National Natural Science Foundation of China under Contracts No.~10575109, No.~10775142, No.~10875115, No.~11175187, and  No.~11475187; the Ministry of Education, Youth and Sports of the Czech Republic under Contract No.~LG14034; the Carl Zeiss Foundation, the Deutsche Forschungsgemeinschaft and the VolkswagenStiftung; the Department of Science and Technology of India; the Istituto Nazionale di Fisica Nucleare of Italy; National Research Foundation (NRF) of Korea Grants No.~2011-0029457, No.~2012-0008143, No.~2012R1A1A2008330, No.~2013R1A1A3007772, No.~2014R1A2A2A01005286, No.~2014R1A2A2A01002734, No.~2014R1A1A2006456; the Basic Research Lab program under NRF Grant No.~KRF-2011-0020333, No.~KRF-2011-0021196, Center for Korean J-PARC Users, No.~NRF-2013K1A3A7A06056592; 
the Brain Korea 21-Plus program and the Global Science Experimental Data Hub Center of the Korea Institute of Science and Technology Information; the Polish Ministry of Science and Higher Education and the National Science Center; the Ministry of Education and Science of the Russian Federation and the Russian Foundation for Basic Research; the Slovenian Research Agency;
the Basque Foundation for Science (IKERBASQUE) and the Euskal Herriko Unibertsitatea (UPV/EHU) under program UFI 11/55 (Spain);
the Swiss National Science Foundation; the National Science Council and the Ministry of Education of Taiwan; and the U.S.\
Department of Energy and the National Science Foundation. This work is supported by a Grant-in-Aid from MEXT for Science Research in a Priority Area (``New Development of Flavor Physics'') and from JSPS for Creative Scientific Research (``Evolution of Tau-lepton Physics'').


\begin{thebibliography}{99}
\bibitem{BayesianStat}
In common HEP usage, Bayesian intervals or credibility levels have been reported as ``confidence intervals'' or ``confidence levels,'' which is a frequentist-statistics term.
\bibitem{CC}
Throughout this Letter, the inclusion of the charge-conjugate decay mode is implied.
\end{thebibliography}
\end{document}